\documentclass[aps,prl,twocolumn,showpacs,superscriptaddress,groupedaddress]{revtex4}  

\usepackage{amsmath}
\usepackage{amsfonts}
\usepackage{amssymb}
\usepackage{bbm}
\usepackage{dsfont}

\usepackage{graphicx}
\usepackage{slashed}
\usepackage{color}
\usepackage{hyperref}
\usepackage{booktabs}

\newcommand{\imag}{\text{i}}
\newcommand{\skipthis}[1]{}

\usepackage[caption=false]{subfig}
\usepackage[T1]{fontenc}
\usepackage[latin9]{inputenc}
\usepackage{amsmath,amssymb,amsthm,amsfonts}
\usepackage{ mathrsfs }
\usepackage{slashed}
\usepackage{graphicx}
\usepackage{color,rotating}
\usepackage{dsfont}
\usepackage{setspace}
\usepackage{verbatim}
\usepackage{fancyhdr}
\usepackage{hyperref}

\def\s0#1#2{\mbox{\small{$ \frac{#1}{#2} $}}}
\def\0#1#2{\frac{#1}{#2}}

\newcommand{\beq}{\begin{equation}}
\newcommand{\eeq}{\end{equation}}
\newcommand{\beqa}{\begin{eqnarray}}
\newcommand{\eeqa}{\end{eqnarray}}

\graphicspath{{./figures/}}

\newcommand{\bea}{\begin{eqnarray}}
\newcommand{\eea}{\end{eqnarray}}

\definecolor{darkgreen}{rgb}{0,0.6,0}
\definecolor{gray}{rgb}{.7,.7,.7}

\def\eq#1{(\ref{#1})}

\newcommand {\apgt} {\ {\raise-.5ex\hbox{$\buildrel>\over\sim$}}\ }
\newcommand {\aplt} {\ {\raise-.5ex\hbox{$\buildrel<\over\sim$}}\ }

\def\s0#1#2{\mbox{\small{$ \frac{#1}{#2} $}}}
\def\0#1#2{\frac{#1}{#2}}

\newcommand{\be}{\begin{eqnarray}}
\newcommand{\ee}{\end{eqnarray}}

\begin{document}

\author{Nicolai Christiansen}
\affiliation{Institut f\"ur Theoretische Physik, Universit\"at Heidelberg,
Philosophenweg 16, 69120 Heidelberg, Germany}
\author{Michael Haas}
\affiliation{Institut f\"ur Theoretische Physik, Universit\"at Heidelberg,
Philosophenweg 16, 69120 Heidelberg, Germany}
\author{Jan M. Pawlowski}
\affiliation{Institut f\"ur Theoretische Physik, Universit\"at Heidelberg,
Philosophenweg 16, 69120 Heidelberg, Germany}
\affiliation{ExtreMe Matter Institute EMMI, GSI Helmholtzzentrum f\"ur
Schwerionenforschung mbH, Planckstr.\ 1, 64291 Darmstadt, Germany}
\author{Nils Strodthoff}
\affiliation{Institut f\"ur Theoretische Physik, Universit\"at Heidelberg,
Philosophenweg 16, 69120 Heidelberg, Germany}


\title{Transport Coefficients in Yang--Mills Theory and QCD}

\begin{abstract}
We calculate the shear viscosity over entropy density ratio $\eta/s$ in
Yang--Mills theory from the Kubo formula using an exact diagrammatic
representation in terms of full propagators and vertices using gluon spectral
functions as external input. 
We provide an analytic fit formula for the temperature dependence of $\eta/s$ over the whole temperature range
from a glueball resonance gas at low temperatures, to a high-temperature regime
consistent with perturbative results. 
Subsequently we provide a first estimate for $\eta/s$ in QCD.
\end{abstract}

\maketitle
{\it Introduction - } The experimental heavy-ion programs at
RHIC \cite{Adcox:2004mh,Adams:2005dq} and at the LHC \cite{Aamodt:2010pa}
explore the physics of the quark-gluon plasma (QGP). 
It turns out that the dynamics of the hot plasma
created in heavy-ion collisions is well-described by
hydrodynamics. Therefore, the determination of transport coefficients in the
QGP is of great interest. One aspect is that the inference of the initial state
physics requires a precise description of the hydrodynamical evolution, which
in turn depends on transport coefficients as microscopic input,
\cite{Niemi:2011ix}. In particular, the viscosity over entropy ratio $\eta/s$
governs the
efficiency of the conversion of the initial spatial anisotropy into a momentum
anisotropy of the final state. 

For the determination of $\eta/s$ and its temperature dependence in
the quark-gluon plasma, theoretical approaches face several
challenges. The temperature regimes below and above the critical
temperature $T_c$ are characterised by different degrees of freedom,
and for temperatures $T \lesssim 2 T_c$ non-perturbative effects
become important. Of particular interest is the vicinity of $T_c$,
where the minimum for $\eta/s$ is expected
\cite{Csernai:2006zz,Hirano:2005wx}.  A universal lower bound for
$\eta/s$ of $1/4\pi$ was conjectured in \cite{Kovtun:2004de} using the
AdS/CFT correspondence. Indeed, measurements of the elliptic flow
$v_2$ indicate a value for $\eta/s$ which is of the order of this
lower bound \cite{Heinz:2013th}. The bound has been tested
theoretically with several methods for the QGP
\cite{Haas:2013hpa,Aarts:2002cc,Xu:2007ns,Meyer:2007ic,Meyer:2009jp,
Lang:2013lla,Marty:2013ita}, but also for other potentially perfect liquids,
such
as ultracold atoms \cite{Gelman:2004fj,Thomas:2009,Schaefer:2014awa}.
 
The Kubo formulae relate $\eta$ to the energy-momentum tensor (EMT)
\cite{Kubo:1957mj}. Spectral functions are real-time quantities and
cannot be obtained directly from Euclidean correlation functions.
However, the direct calculation of real-time correlation functions
represents a notoriously difficult problem in non-perturbative
approaches to quantum field theory. Even though first computations in
this direction have been performed e.g. in
\cite{Kamikado:2013sia,Strauss:2012dg}, we shall
utilise Euclidean correlation functions within a numerical analytic
continuation.

In this work we study the shear viscosity over entropy ratio $\eta/s$
in pure $SU(3)$ Landau gauge Yang-Mills (YM) theory within the
approach set-up in \cite{Haas:2013hpa}. In the present work we
considerably generalise the approach, also aiming at quantitative
precision. We apply an exact functional relation that allows a
representation of the EMT correlation function in terms of full
propagators and vertices of the gluon field. The analysis covers the
entire temperature range from the glueball regime below the critical
temperature $T_c$, up to the ultraviolet where perturbation theory is
applicable. In particular this resolves the non-perturbative domain at
temperatures $T \lesssim 2 T_c$.  We provide a global, analytic fit
formula for $\eta/s$ which extends the well-known perturbative
high-temperature behaviour to the non-perturbative temperature regime.
Based on this description for pure gauge theory, a first
estimate for $\eta/s$ in full QCD is derived.

\begin{figure}[b]
\centering
\includegraphics[width=0.8\columnwidth]{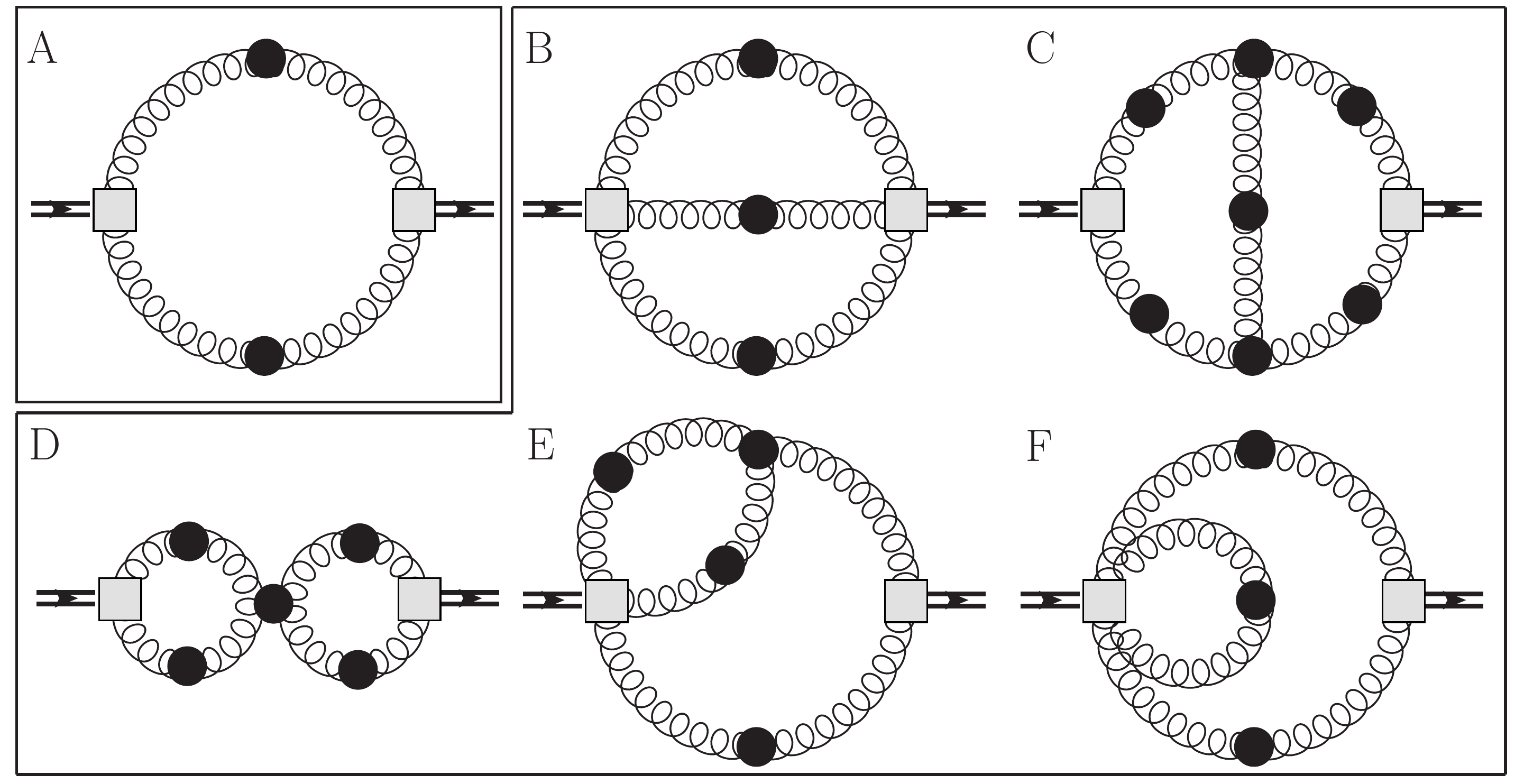}
\caption{Types of diagrams contributing to the correlation function of the
energy momentum tensor up to two-loop order; squares denote vertices derived
from the EMT; 
all propagators and vertices are fully dressed.}
\label{fig:transport2loop} 
\end{figure}

{\it YM shear viscosity from gluon spectral functions - } The shear
viscosity is related to the spectral function $\rho_{\pi\pi}$ of the
spatial traceless part $\pi_{ij}$ of the energy momentum tensor tensor
via the Kubo relation
\begin{equation}
\label{eq:Kubo}
\eta=\lim\limits_{\omega \to 0} \0{1}{20}\frac
{\rho_{\pi\pi}(\omega,\vec{0}\,)}{\,\omega}\,,
\end{equation}
where
\begin{equation}
\label{eq:EMT}
\rho_{\pi\pi}(\omega,\vec{p}\,)= \int\hspace{-2mm} \tfrac {d^4 x} {(2\pi)^4}  
e^{-i\omega x_0+i\vec{p}\vec{x}}\langle[\pi_{ij}(x),\pi_{ij}(0)]\rangle\,.
\end{equation}
For the computation of \eq{eq:EMT} we use the fact that a general correlation
function of composite operators can be expanded in terms of full propagators
and full vertices of
the elementary fields \cite{Haas:2013hpa,Pawlowski:2005xe},
\begin{equation}\label{eq:magicformula}
  \langle \pi_{ij}[\hat A] \pi_{ij}[\hat A]\rangle = \pi_{ij}[ G_{A
\phi_k}\!\cdot\! 
  \tfrac{\delta}{\delta \phi_k}
  +A]\, \pi_{ij}[ G_{A \phi_k}\!\cdot\! \tfrac{\delta}{\delta
\phi_k} + A]\,, 
\end{equation}
where $\phi=(A,c,\bar c)$ denotes the expectation value of
the fluctuation (super-)field $\hat \phi$, e.g.\ $A=\langle \hat A\rangle$, and
$G_{\phi_i \phi_j}=
\langle \hat \phi_i \, \hat \phi_j\rangle - \langle \hat \phi_i\rangle \,\langle
\hat \phi_j\rangle$
denotes the propagator of the respective fields. This yields a diagrammatic 
representation in terms of a finite number of diagrams involving full
propagators
and vertices, see Fig.~\ref{fig:transport2loop} for the types of diagrams appearing
in the full expansion up to two-loop order. We emphasise that
\eq{eq:magicformula} is
an exact relation whose finite diagrammatics should not be confused with a
perturbative
expansion in an infinite series of Feynman diagrams. 
The internal vertices arise from functional derivatives of the full propagator
in \eq{eq:magicformula} and are therefore automatically fully dressed.
However, the RG-invariance of the left hand side of \eq{eq:magicformula} only carries
over to right hand side if also the external vertices derived from the EMT are dressed
with appropriate wave-function renormalisation factors and running couplings.
This argument
is supported by the flow equation for the EMT itself, which can be derived 
from the flow equation for composite operators \cite{Pawlowski:2005xe},
where full vertices are generated during the flow.
More heuristically this can also be seen in a skeleton expansion. Therefore, on
a diagrammatic level only up to 3-loop diagrams with dressed external
vertices appear.

The natural framework for such a calculation is the
real-time formalism based on the Schwinger-Keldysh closed time path. Within
such a setup one never has to resort to Euclidean field theory.
Here one distinguishes two branches of the time contour, conventionally denoted
by $+/-$, along with separate fields and sources. Correlation functions thus
become matrix valued. In thermal equilibrium the
propagator can be parametrised in terms of the spectral function
$\rho(\omega,\vec{p})$ only according to
\begin{eqnarray} 
 G^{\pm\pm}(\omega,\vec{p})&=&F(\omega,\vec{p}) \pm \imag
\left(n(\omega)+\tfrac{1}{2}\right)\rho(\omega,\vec{p})\nonumber\,,\\
 G^{+-}(\omega,\vec{p})&=&-\imag \, n(\omega)\rho(\omega,\vec{p})\nonumber\,,\\
 G^{-+}(\omega,\vec{p})&=&-\imag
\left(n(\omega)+1\right)\rho(\omega,\vec{p}) \,,
\label{eq:Gs}\end{eqnarray}
where $n(\omega)=1/(\exp(\omega/T)+1)$ denotes the Bose distribution function and $F(\omega,\vec{p})$ is given as a
principal value integral,
\begin{equation}
F(\omega,\vec{p})=PV
\int\limits_{-\infty}^\infty\hspace{-2mm}\text{d} \bar\omega\, \frac{\rho(\bar{\omega},\vec{p})}{\omega-\bar{\omega
}}\,.
\end{equation}
The spectral function is defined as
\begin{equation}
\rho( \omega,\vec{p} ) = G^{-+} ( \omega,\vec{p} ) - G^{+-}( \omega,\vec{p} )
\,.
\end{equation}
Moreover, in thermal equilibrium the KMS relation relates the off-diagonal parts of the propagator via
\begin{equation}
G^{+-}(\omega,\vec{p})=e^{-\beta\omega}G^{-+}(\omega,\vec{p}) \,.
\end{equation}
Hence we find for the spectral function of the energy momentum tensor
\begin{equation}
\label{eq:EMtspectralfunctionRTF}
\rho_{\pi\pi}(\omega,\vec{p})=(1-e^{-\beta\omega})G^{-+}_{\pi\pi}(\omega,\vec{p}) \,.
\end{equation}
Inserting the above identity into \eq{eq:Kubo}, this implies
\begin{equation}
\label{eq:eta}
\eta=-\frac \beta {20} G^{-+}_{\pi\pi}(0,0) \,.
\end{equation}

In this work we present the full two-loop diagrammatics shown in
Fig.~\ref{fig:transport2loop}. 
There are five types of two-loop diagrams arising from the expansion
(\ref{eq:magicformula}): Sunset (B), Maki-Thompson (C), Eight (D),
Squint (E), one-loop with vertex correction (F).  The branch
indices of the external vertices are fixed by \eq{eq:eta} as $-+$,
whereas we sum over internal branch indices. Thus, unlike in the
one-loop case, at two-loop level principal value parts of propagators with equal
branch indices
can occur. However, at two-loop level possibly divergent contributions can explicitly be shown
to cancel due to a left-right symmetry after combining
appropriate diagrams.  This is no longer true beyond two-loop, where
diagrams with divergent sub-diagrams arise.

The only nontrivial input in our calculation, apart from the running
coupling $\alpha_s$, is the gluon spectral function obtained using MEM
from Euclidean FRG data \cite{Fister:2011uw}.  For details about MEM
and the properties of the gluon spectral functions we refer the reader
to \cite{Haas:2013hpa}. The running coupling $\alpha_s(q,T)$ extracted
from the ghost-gluon vertex is calculated from the dressing functions
$z_{\bar{c}Ac},Z_c,Z_T$ of the ghost-gluon-vertex, the ghost
propagator and the transverse gluon propagator, respectively as
\begin{equation}
\label{eq:alphasvertex}
 \alpha_s(q,T)=\frac{z_{\bar{c}Ac}^2(q,T)}{4\pi Z_T(q,T)Z_c(q,T)^2}
\end{equation}
with data taken from \cite{Fister:2011uw}. Following the discussion above, all 
couplings that appear in the vertices are fully dressed running couplings.
For each two-loop diagram we study the integrand of the viscosity
integral as a function of one of the loop four-momenta $(q_0,\vec{q})$,
integrating out the other one. It turns out that all integrands
are peaked in the vicinity of some diagram-dependent value
$(q_{0,\rm{max}},\vec{q}_{\rm{max}})$.  The running couplings
$\alpha_s(q,T)$ are then evaluated at a momentum
$q_{\mathrm{max}}(T)=\sqrt{q_{0,\rm{max}}^2+\vec{q}_{\rm{max}}^2}\approx 7\, T$
to minimise
the impact of the neglected momentum dependence of the vertices. This implicitly defines
a temperature-dependent vertex coupling
$\alpha_{s,\text{vert}}(T)=\alpha_s(7\,T,T)$.

\begin{figure}[t]
\centering
\includegraphics[width=\columnwidth]{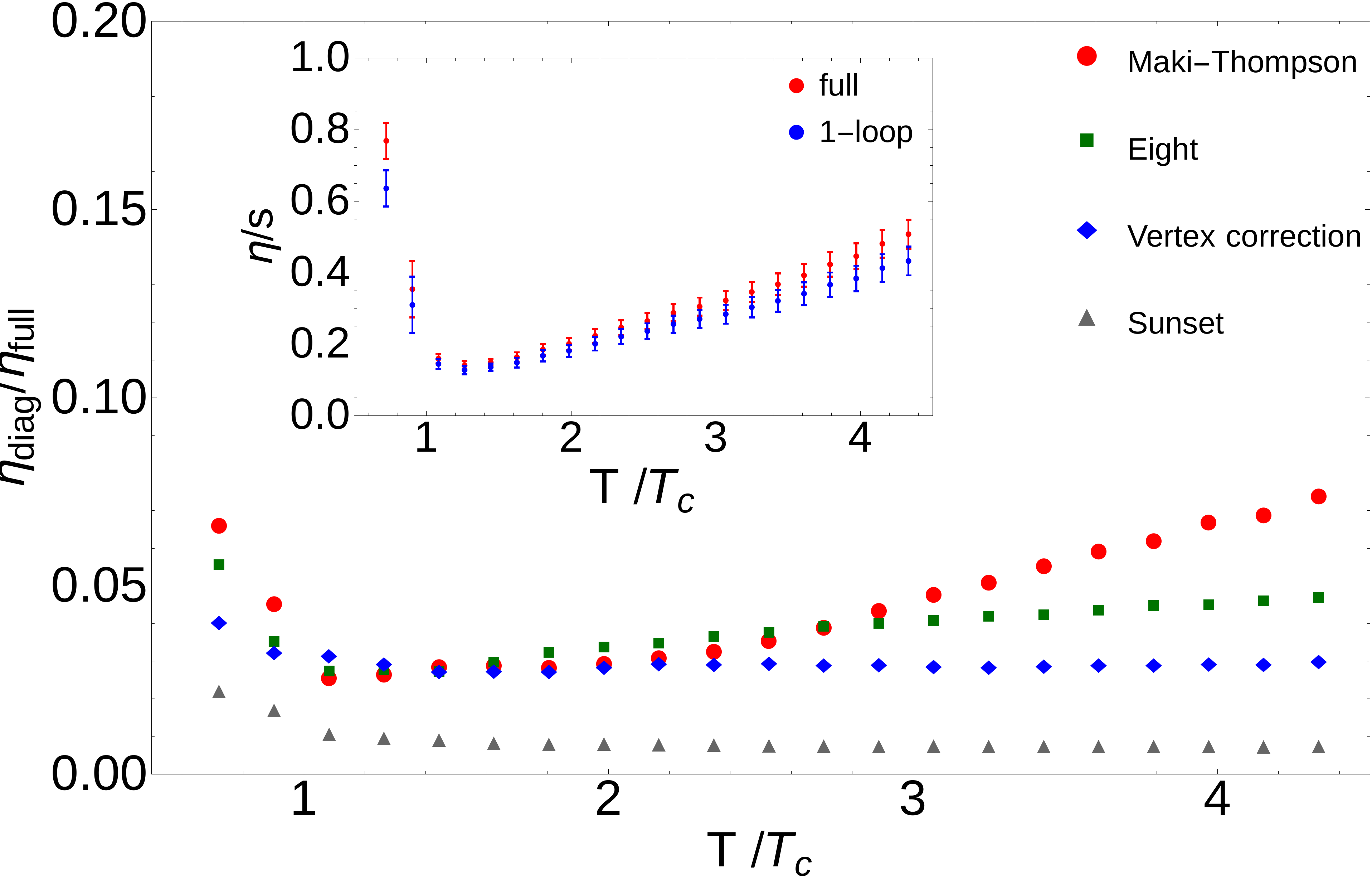}
\caption{Relative contributions from different diagram types to the
  two-loop viscosity as a function of temperature. The squint
  contribution is orders of magnitude smaller and not shown.  The
  inset shows the comparison to the one-loop result
  \cite{Haas:2013hpa}.}
\label{fig:voe_two_loop_contributions} 
\end{figure}

{\it Results - }
\begin{figure}[b]
\centering
\includegraphics[width=\columnwidth]{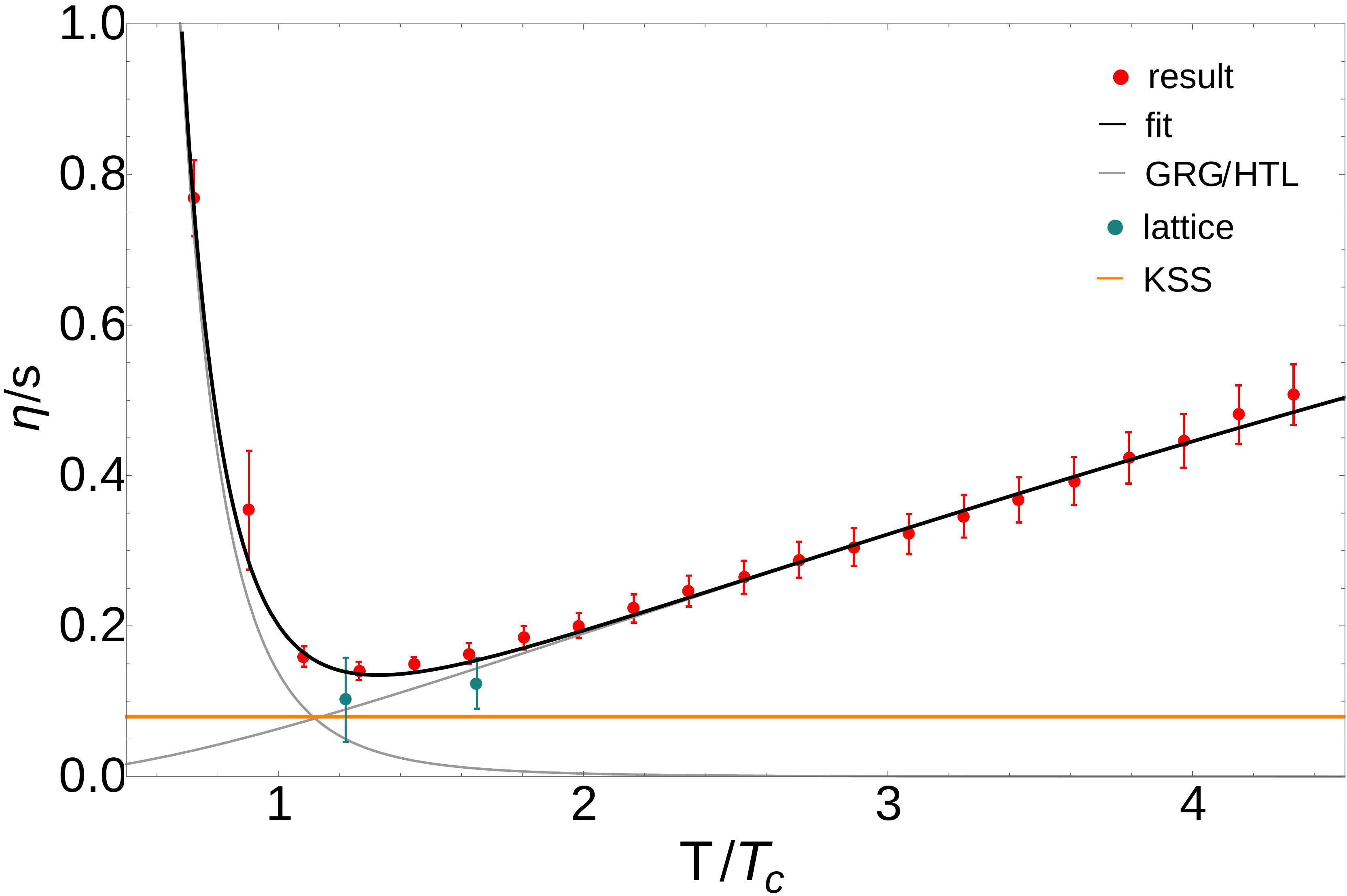}
\caption{Full Yang-Mills result (red) for $\eta/s$ in comparison to lattice
results \cite{Meyer:2007ic,Meyer:2009jp} (blue) and the AdS/CFT bound (orange).
In addition, the plot shows the 
analytic fit given in \eq{eq:globalYM} and its two components. The ratio
$\eta/s$ shows a minimum at $T_{\mathrm{min}} \approx 1.26 \, T_c$ with a value
of
0.14.} 
\label{fig:voe_twoloop} 
\end{figure}
Fig.~\ref{fig:voe_twoloop} shows the full two-loop result for $\eta/s$ employing
the lattice entropy density from \cite{Borsanyi:2012ve}
including all diagrams from Fig.~\ref{fig:transport2loop}. 
The data shows, as expected on general grounds, a clear minimum at
$T_{\mathrm{min}} \approx 1.26 \, T_c$. The minimal value
$\eta/s(T_{\mathrm{min}})=0.14$ is well above the AdS/CFT bound, where
the error bars represent the combined systematic
errors from MEM and the FRG calculation. 
The lattice data \cite{Meyer:2007ic,Meyer:2009jp} is in good agreement with our
results, supporting the reliability of both methods. 
The inset in Fig.~\ref{fig:voe_two_loop_contributions} shows the
comparison to the one-loop calculation \cite{Haas:2013hpa}, illustrating the
very good
agreement around $T_c$. This confirms 
the argument concerning the optimisation of the RG scheme around $T_c$,
which was put forward in \cite{Haas:2013hpa}. 
Consistent with this reasoning, 
only at larger temperatures the deviation between the two calculations
becomes significant and the relative size of the two-loop contribution
grows with temperature. 
For large temperatures the
dominant two-loop contributions arise from the Maki-Thompson and the Eight, see
Fig.~\ref{fig:voe_two_loop_contributions}, that resum classes of
ladder diagrams. This is consistent
with the conventional picture in perturbative expansions where ladder
resummations are required to obtain the correct result for the
viscosity \cite{Jeon:1994if,Jeon:1995zm}. Note that diagrams with overlapping
loops are potentially suppressed as the spectral functions are peaked in a
narrow region in momentum space. Due to the additional phase space
suppression, we expect that diagrams with more than two loops are negligible. We
have checked this suppression in a first assessment of three-loop diagrams.

For understanding the physical picture underlying the temperature
behaviour, we provide a global fit function for $\eta/s \, (T)$.
Additionally, such an analytic fit function is well-suited for
phenomenological applications.  This parametrisation has to cover
temperature ranges corresponding to vastly different physical
situations. At large temperatures $T\gg T_c$ the degrees of freedom
are gluons which can eventually be treated perturbatively.  By
contrast, at small temperatures $T\lesssim T_c$ YM theory can
effectively be described as a glueball resonance gas (GRG). Finally,
there is a transition region between these two asymptotic regimes
whose description requires non-perturbative techniques.

In the high temperature regime, perturbation theory is applicable and
$\eta/s$ is given as a function of the strong coupling $\alpha_s$ only. It turns
out that the hard-thermal loop (HTL) resummed data \cite{Arnold:2003zc} is
well-described by the functional form
\begin{equation}
\frac{\eta}{s}(\alpha_s) = \frac{a}{\alpha_s^{\gamma}} \, , 
\label{eq:alpha_scaling}
\end{equation}
with an overall coefficient $a$ and a scaling exponent $\gamma\approx
1.6$. We aim at extracting a non-perturbative extension of the above
parametrisation based on our data. In the region $T_c-3\,T_c$ strong
correlations become important and perturbation theory breaks
down. This raises the question of a suitable running coupling as there
is no unique definition of $\alpha_s$ beyond two-loop. A
quasi-particle picture suggests that an appropriate choice of
$\alpha_s$ can be deduced from a heavy quark potential
\cite{Richardson:1978bt,Karbstein:2014bsa}. 

An analytic expression for a coupling that generates a linearly rising
static quark potential at large distances is given by
\cite{Nesterenko:1999np}
\begin{equation}
\label{eq:alphasheavyquark}
\alpha_{\mbox{\tiny{s,HQ}}}(z)=\frac{1}{\beta_0}\frac{z^2-1}{ z^2 \log z^2}\,,
\end{equation}
where $z$ denotes a dimensionless momentum variable.  At large momenta
it approaches the one-loop running coupling, where
$\beta_{0}=33/(12\pi)$ denotes the coefficient in the one-loop
beta-function of pure $SU(3)$ Yang-Mills theory. The scale
identification is implemented by regarding
$\alpha_{\mbox{\tiny{s,HQ}}}$ as a function of $z=c \, T/T_c$ with a
scale identification factor $c$. By construction, the divergence of
\eq{eq:alphasheavyquark} at zero momentum leads to a vanishing
contribution of \eq{eq:alpha_scaling} to $\eta/s$ at zero
temperature. As an estimate for a lower bound for a reasonable
high-temperature fit, we consider the trace anomaly as a hint from QCD
thermodynamics, which starts to develop a $T^4$ behaviour for
$T\gtrsim 2\, T_c$ \cite{Andersen:2011ug}. Using $T>3\, T_c$ as a
conservative estimate, our data is well-described by the scaling form
\eq{eq:alpha_scaling} with the running coupling
\eq{eq:alphasheavyquark} and parameters $a=0.15$ and $c = 0.66$. One
should note that whereas the heavy quark potential coupling takes a
rather large value $\alpha_{\mbox{\tiny{s,HQ}}}(c
T/T_c)|_{T=T_c}\approx 1.77$ at $T_c$, the vertex coupling
$\alpha_{\mbox{\tiny{s,vert}}}(T_c)\approx 0.76$ corresponding to a
value of
$\alpha^{\overline{\text{MS}}}_{\mbox{\tiny{s,vert}}}(T_c)\approx
0.35$, after conversion to the $\overline{\text{MS}}$ scheme
\cite{vonSmekal:2009ae}, is comparably small. This supports the
validity of resummation arguments at moderately large temperatures but
also underlines the non-uniqueness of the definition of a running
coupling in the nonperturbative regime around $T_c$.  It turns out
that the fit \eq{eq:alpha_scaling} can be extended to even lower
temperatures $T\gtrsim 1.8 T_c$, where it is still in very good
agreement with our data, see Fig.~\ref{fig:voe_twoloop}. Note, that
the fitting with the vertex coupling $\alpha_{s,\text{vert}}$ fails
for temperatures below $3\, T_c$. These findings hint at the validity of a
quasi-particle picture even at considerably low temperatures.

Below the critical temperature the effective degrees of freedom change
from gluons to glueballs. The glueball spectrum can be calculated
using the formalism put forward in this work \cite{HMP}. Hence, the
present YM calculation is also capable of describing glueball
resonances. Therefore one expects an algebraic decay of $\eta/s$ with
temperature similar to a hadron resonance gas
\cite{Csernai:2006zz,Hirano:2005wx}. Due to the small number of data
points and the comparably large error bars below $T_c$, no precise
determination of the exponent $\delta$ in the
power law is possible. We construct a global fit function by superposing a
power law behaviour at small temperatures with the extrapolated high temperature
behaviour \eq{eq:alpha_scaling}, i.e.\ a global parametrisation of the form
\begin{equation}
\label{eq:globalYM}
\frac{\eta}{s}(T) =
\frac{a}{\alpha_{\mbox{\tiny{s,HQ}}}^{\gamma}(c\,T/T_c)}
 +
\frac{b}{(T/T_{c})^{\delta}} \,.
\end{equation}
With $a=0.15$, $b=0.14$, $c = 0.66$ and
$\delta=5.1$ this fit describes our data very
well, see Fig.~\ref{fig:voe_twoloop}. The best-fit value
$\delta=5.1$ lies in the expected range for a hadron
resonance gas \cite{Hirano:2005wx}, where for example a pion gas leads to an
exponent of 4.

\begin{figure}[t]
\centering
\includegraphics[width=\columnwidth]{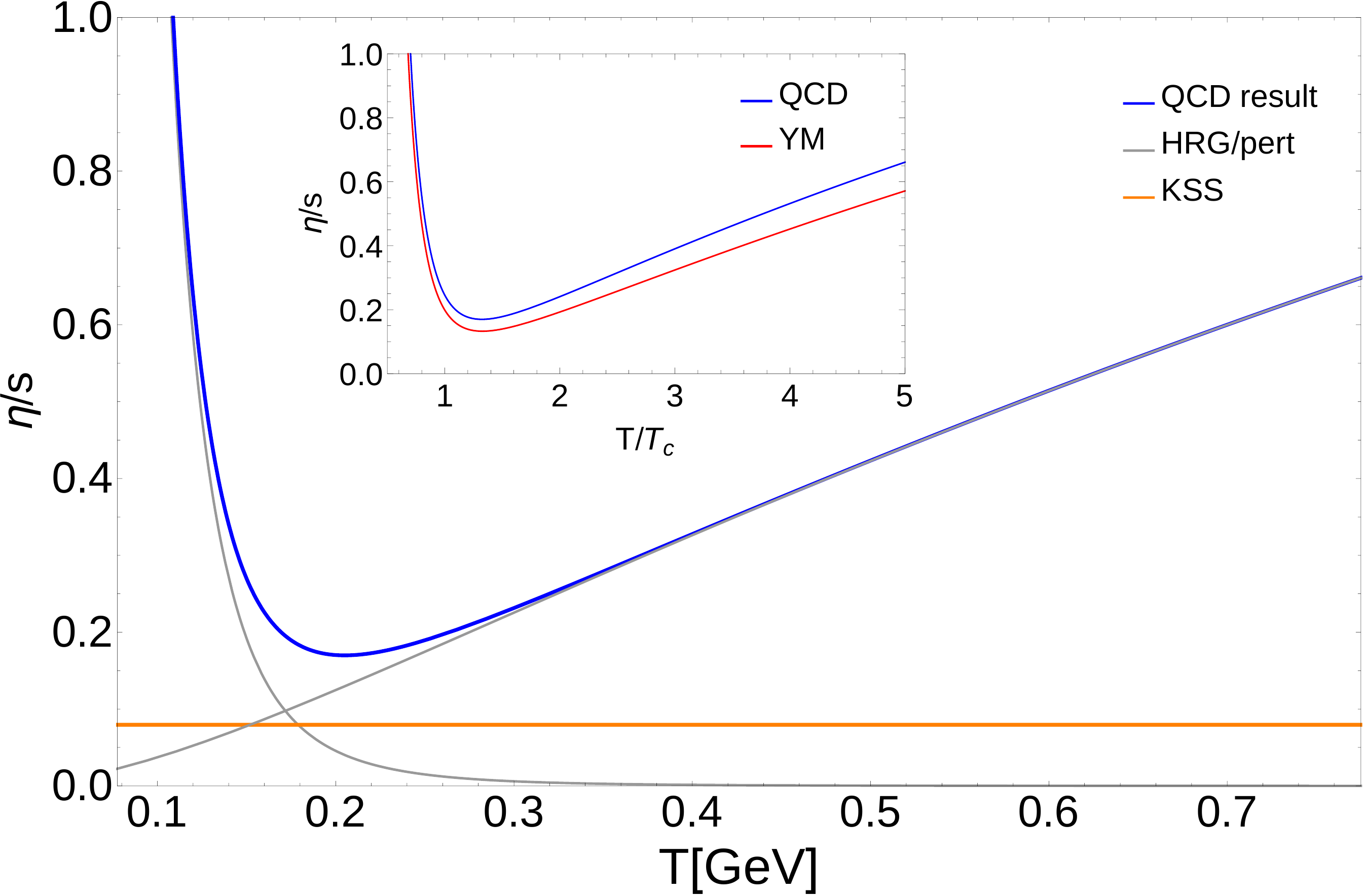}
\caption{Estimate for $\eta/s$ in QCD which shows a minimum at
  $T_{\mathrm{min}} \approx 1.3 \, T_c$ at a value of $0.17$. The inset shows
the
  comparison to the YM results for temperatures normalised by the
  respective critical temperatures.}
\label{fig:voe_qcd} 
\end{figure}

The analytic fit function \eq{eq:globalYM} for $\eta/s$ in YM theory enables us to
provide a first estimate of $\eta/s$ in full QCD, again based on the idea
of superposing a low and a high temperature behaviour term.
The procedure consists of three separate steps.  
Firstly, one has to take into account the difference in scales
and the running couplings in YM and QCD.
This involves replacing the coefficient $\beta_0$ in \eq{eq:alphasheavyquark} by
its QCD value, $\beta_{0,\text{\tiny{QCD}}}=(33-2 N_f)/(12\pi)$.
Additionally, one has to set a scale by fixing the ratio of the running
couplings in YM and QCD at a certain point. In our setup the characteristic
scale
is the critical temperature $T_c$. For the phase
transition to the confinement phase to take place, the strong coupling usually
needs to exceed a certain critical value
$\alpha_s(T)=\alpha_{\mathrm{crit}}$. On general grounds one can argue that the
critical values in YM theory and QCD are of comparable size. 
This argument is supported by the fact that the values of
$\alpha_{\mathrm{crit}}$ for the vertex couplings tend to coincide.
Consequently, we impose the condition
\begin{equation}
\alpha^{N_f=0}_{\mbox{\tiny{s,HQ}}}(cT/T_c)\Bigr|_{T=T_{c}}=\alpha^{N_f=3}_{
\mbox{\tiny{s,HQ}}}(c_{\text{\tiny QCD}}T/T_c)\Bigr|_{T=T_{c}}.
\end{equation}
This matching condition fixes the scale factor to the value $c_{\text{\tiny
QCD}}= 0.79$.
Secondly, one has to take into account genuine quark contributions that are not
encoded
in the change of the running couplings. Denoting the quark contributions to
viscosity and entropy as 
$\Delta \eta$ and $\Delta s$ respectively, we write
\begin{equation}
\begin{split}
  \frac{\eta}{s}\Bigr|_\text{\tiny{QCD}}&=\frac{\eta_\text{\tiny{YM}}+\Delta
    \eta}{s_{\text{\tiny{YM}}}+\Delta
    s}=\frac{\eta}{s}\Bigr|_{\text{\tiny{YM}},
\alpha^{\text{\tiny YM}}_s\to\alpha^{\text{\tiny QCD}}_s
}\cdot\left(\frac{1+\frac{\Delta
        \eta}{\eta^{\phantom{YM}}_\text{YM}}}{1+\frac{\Delta
        s}{s^{\phantom{YM}}_\text{YM}}}\right),
\end{split}
\end{equation}
and estimate the ratios $\Delta \eta/\eta_\text{YM}$ and $\Delta s/s_\text{YM}$ 
using leading order perturbative results. For $N_f=3$ we find $\Delta
\eta/\eta_\text{YM}\approx 2.9$ \cite{Chen:2012jc,Arnold:2000dr} and
$\Delta s/s_\text{YM}\approx \frac{21}{32} N_f\approx 2.0$
\cite{Arnold:1994eb,Andersen:1999va}, leading to an overall correction factor
of approximately $4/3$. Finally, in the low temperature regime one has to
replace the
pure glueball resonance gas by a hadron resonance gas, which also decays algebraically
with temperature. In this work we use the data given in
\cite{Demir:2008tr}.
In summary, the final fit for QCD takes the form \eq{eq:globalYM}, but with the
parameter $a_{\text{\tiny QCD}} \approx 4/3 \, a$
for the high-temperature part and $b_{\text{\tiny QCD}} \approx 0.16$, 
$\delta_{\text{\tiny QCD}}\approx 5$ for the HRG fit, replacing the
corresponding YM values. Additionally the full QCD $\alpha^{N_f=3}_{
\mbox{\tiny{s,HQ}}}(c_{\text{\tiny QCD}}T/T_c)$ with $c_{\text{\tiny QCD}}=
0.79$ replaces the pure-glue beta-function, whereas the perturbative exponent $\gamma$ remains
unchanged. Note that a continuation of the fit to very high energies requires
taking into account the quark flavor thresholds appropriately.

This procedure yields the final result shown in
Fig.~\ref{fig:voe_qcd}. Plotted in terms of temperatures normalised by the
respective critical temperatures, the QCD curve is shifted slightly upwards
compared to the YM result, see the inset of Fig.~\ref{fig:voe_qcd}. The general
shape resembles the one of the YM result and shows a minimum at
$T_{\mathrm{min}} \approx 1.3 \, T_c$ with a value $0.17$.

{\it Summary and Conclusions - } We have computed the shear viscosity
over entropy density ratio in pure YM theory over a large temperature
range. The setup is based on an exact functional
relation for the spectral function of the
energy-momentum tensor involving full gluon propagators and vertices. The only
input are the gluon spectral function and the running coupling
$\alpha_s$.  As a highly non-trivial result, the global temperature
behaviour of $\eta/s$ can be
described as a direct sum of a glueball resonance gas contribution
with an algebraic decay at small temperatures, and a high temperature
contribution consistent with HTL-resummed perturbation
theory. Finally we provide a first estimate for $\eta/s$ in QCD.

\noindent {\bf Acknowledgements} The authors thank L. McLerran, G. Denicol, 
H. Niemi, and D. Rischke for discussions. 
This work is supported by the Helmholtz Alliance HA216/EMMI and the grant
ERC-AdG-290623. NC acknowledges funding
from the Heidelberg Graduate School of Fundamental Physics.

\bibliographystyle{bibstyle}
\bibliography{bibfile}

\end{document}